\documentclass[twocolumn,aps,prd,preprintnumbers,nofootinbib]{revtex4}
\usepackage[dvips]{graphicx}
\usepackage{bm,latexsym,amsmath,amssymb,amsfonts,color}

\usepackage[normalem]{ulem}
\newcommand{\D}{{\rm d}}

\begin{document} 

\title{\uline{}Modeling scalar fields consistent with positive mass} 

\author{Masato Nozawa${}^{1}$ and Tetsuya Shiromizu${}^{2}$}
\affiliation{${}^{1}$Theory Center, KEK, Tsukuba 305-0801, Japan}
\affiliation{${}^{2}$Department of Physics, Kyoto University, Kyoto 606-8502, Japan}

\begin{abstract}
This paper explores the conditions under which modified gravitational theories 
admit the positive mass. Following Witten's spinor argument,  it is argued that a 
single condition should be imposed upon a gauge connection in the
 super-covariant derivative. Under this condition,  
we present a simple formula for the divergence of the Nester tensor in Einstein's gravity with general source. 
Applying this prescription to the Einstein-scalar system, we find that
 for a certain class of the gauge connection,  
a special kind of non-canonical scalar-field theory admits the positivity property 
in addition to the ordinary canonical scalar-field system.  
In both cases the scalar potential can be written in terms of a superpotential.  
In the non-canonical case we obtain the most general ``BPS'' solutions
 which preserve at least half of the supersymmetry.
\end{abstract}
\maketitle

\section{Introduction}

The observational evidence of the present-day acceleration of the universe~\cite{Riess}
brought us with a tantalizing and profound puzzle in modern cosmology. 
A possible resolution to this problem may be ascribed to the presence of an unknown component of the energy-momentum 
tensor mimicking a positive cosmological constant, dubbed as dark
energy.  A simple instance is provided by the scalar fields, which are ubiquitous in string theory. 
A plethora of scalar-field models of dark energy 
are able to alleviate coincidence problems and fine-tuning. 
Another approach  is to modify general relativity at the cosmological distance. 
However, most of modified theories of gravity which were invented to account for the late-time acceleration of the universe 
are phenomenological and are lack of a fundamental theoretical description.   
Sometimes modified gravities require nonlinear effects to 
recover general relativity, resulting in  poor theoretical predictability.  
Moreover, many of these theories do not respect suitable energy 
conditions.  These properties cast doubt on the reliability of gravitational theories. 
In general relativity, on the other hand, 
the positive mass theorem guarantees the classical 
non-perturbative stability of spacetimes~\cite{Schoen, Witten}, 
and the zero-mass ground state is the flat Minkowski spacetime.  
At the classical level, the positivity of mass undoubtedly places a strong underpinning
for viable gravitational theories, and at the quantum level
the instability of classical theory could cause unitary violation. 
A nontrivial example is Starobinsky's $R+\alpha R^2$
gravity~\cite{Starobinsky:1980te}. 
If $1+2\alpha R>0$ (this condition also assures that the Cauchy problem
is well-posed),
one can perform the conformal transformation of the metric to bring this theory into the scalar-tensor 
theory in which the energy-momentum tensor 
obeys the dominant energy condition. Hence this theory admits the mass 
positivity~\cite{Strominger:1984dn}.

The present paper aims to extend Witten's argument~\cite{Witten}
to gravitational theories incorporating a non-canonical scalar
field. Specifically, 
our discussion will be primarily concentrated on a theory  
described by the Lagrangian $R+2K(\phi, X)$, where $X:=-(1/2)(\nabla\phi)^2$. 
This class of Lagrangian has been intensively studied in a cosmological setting since  
it can drive inflation without a finely constructed scalar potential~\cite{ArmendarizPicon:1999rj}.  
This is the special class of Horndeski's scalar-tensor theory~\cite{deRham2011} which 
is the most general scalar-tensor theory maintaining the equations of motion 2nd-order. 
It was recently realized that Horndeski's theory could be obtained in the decoupling limit of the non-linear massive gravity~\cite{deRham2012}. 
We would like to address the viability of these models 
in the context of classical stability.  
In this paper, we study the minimal condition under which the mass positivity is guaranteed in the 
$R+2K(\phi, X)$ theory. 
This result generalizes Boucher's work \cite{Boucher} in which the conditions for canonical scalar fields 
to possess positive mass were worked out  
(see also Ref.~\cite{Townsend}). We obtain a useful formula for the 
mass positivity that can be used in a broad context.  
Utilizing this formula, we find that in a simple class of the gauge connection, 
the non-trivial form of $K$ consistent with the positive mass 
falls into two families. One is the canonical scalar field with its potential
given in terms of a superpotential, recovering the result in the literature~\cite{Boucher,Townsend}. 
Another class of Lagrangian is the non-canonical theory, on which 
we shall focus in the body of text. 
We obtain the general 
Bogomol'nyi-Prasad-Sommerfield 
(BPS) metric obeying the 1st-order differential
equations in the non-canonical case.

The rest of this paper is organized as follows. In
section~\ref{sec:positivemass}, 
we provide the outline of the proof of the positive 
mass theorem in general gravitational theories and 
present a useful formula with wide applicability.  
Section~\ref{sec:Eisnteinscalar} is devoted to examine the Einstein 
gravity with a non-canonical scalar field and to explore 
which types of theories are compatible with the positive mass theorem. 
Section~\ref{sec:SUSY} addresses some issues when the BPS inequality is
saturated. We discuss the multiple scalar generalization in 
section~\ref{sec:multi}. 
The final section summarizes our work and illustrates the recipe of the 
extension to more general cases.

We adopt the convention with mostly-plus metric signature. 
The Clifford algebra is $\{\gamma_a, \gamma_b\}=2g_{ab}$ and 
the Riemann curvature is given by 
$2\nabla_{[\rho }\nabla_{\sigma ]}V^\mu =R^\mu{}_{\nu\rho\sigma}V^\nu $. 
Greek indices ($\mu, \nu,...$) refer to the spacetime component, whereas 
the Latin indices ($a,b,...$) to the frame indices.  
We work in units of $c=8\pi G=1$ throughout the paper.

\section{Positive mass theorem}

\label{sec:positivemass}

Although we will focus on a simple case later, 
we shall not specify the gravitational theory  in this section and  give
a versatile formula in the proof of the mass positivity.  
We attempt to keep the discussion as general as possible.  
Nonetheless, 
the final upshot is quite simple and universal,  and has a potentially wide applicability.    
We follow Witten's spinor argument~\cite{Witten}, which 
provides a covariant integral expression for gravitational 
energy-momentum. 

Let us consider the $d$-dimensional Lorentzian spacetime ($M, g_{\mu\nu}$) 
admitting a spin structure. In terms of a (commuting) Dirac spinor $\epsilon $, 
we define the Nester tensor as~\cite{Nester}
%
\begin{eqnarray}
N^{\mu\nu}:=-i \Bigl( \bar \epsilon \gamma^{\mu\nu\rho} \hat \nabla_\rho \epsilon
-{\overline {{\hat \nabla}_\rho \epsilon}}\gamma^{\mu\nu\rho}\epsilon \Bigr) \,,
\end{eqnarray}
%
where $\gamma^{\mu\nu\rho}=\gamma^{[\mu}\gamma^\nu \gamma^{\rho]}$
and ${\hat \nabla}_\mu$ is the super-covariant derivative operator defined by 
%
\begin{eqnarray}
{\hat \nabla}_\mu \epsilon=\left(\nabla_\mu + {\cal
 A}_\mu\right)\epsilon \,. 
\label{supercovariantderv}
\end{eqnarray}
%
Here $\nabla_\mu $ is an ordinary Levi-Civit\`a covariant derivative and 
${\cal A}_\mu$ is the connection of the spinor bundle taking values in 
${\rm GL}(2^{[d/2]}, \mathbb C)$. 
The Dirac conjugate of a spinor is given by
$\bar \epsilon=i \epsilon^\dagger \gamma^0$, 
hence we have 
\begin{eqnarray}
{\overline {{\hat \nabla}_\mu \epsilon}}
=  i({\hat \nabla}_\mu \epsilon)^\dagger \gamma^0
=  {\overline {\nabla_\mu \epsilon}} -\bar \epsilon \gamma^0 {\cal A}_\mu^\dagger \gamma^0 
= {\overline {\nabla_\mu \epsilon}} -\bar \epsilon {\bar {\cal A}}_\mu \,,
\end{eqnarray}
%
where ${\bar {\cal A}}_\mu:=\gamma^0 {\cal A}_\mu^\dagger \gamma^0$.
The index $0$ stands for the time component in the local vielbein.

Let $\Sigma $ denote the $(d-1)$-dimensional spacelike partial Cauchy
surface with  boundary $\partial \Sigma $ at infinity. 
From the Stokes's theorem, we obtain 
\begin{align}
-\int _\Sigma \nabla_\nu N^{\mu \nu }u_\mu\D \Sigma =  \frac{1}{2}
\int _{\partial \Sigma } N_{\mu\nu }\D S^{\mu \nu } \,,
\label{Stokes}
\end{align}
%
where $u_\mu$ is the future-pointing unit normal to $\Sigma$.
For the asymptotically flat/anti-deSitter spacetimes, 
the right-hand side of the above equation gives rise to 
the globally conserved energy-momentum (contracted by a 
generator of asymptotic symmetry) with the appropriate form 
of $\cal A_\mu $ and with a required fall-off~\cite{Gibbons:1983aq}. 
Here we suppose that the right-hand side of (\ref{Stokes}) 
yields the finite energy-momentum for a given $\mathcal A_\mu $. 
If we are able to show the positivity of the left-hand side, 
we can then prove that the spacetime energy-momentum is timelike, i.e., 
the positive mass follows.

We turn to evaluate the left-hand side of Eq.~(\ref{Stokes}). 
First of all, we notice the following identity 
%
\begin{eqnarray}
\hat \nabla_{[\mu}\hat \nabla_{\nu]}\epsilon=\Bigl(\frac{1}{8}R_{\mu\nu\alpha\beta}\gamma^{\alpha\beta}
+\frac{1}{2}{\cal F}_{\mu\nu}  \Bigr)\epsilon \,,  
\label{Curvatureid}
\end{eqnarray}
%
where $\cal F_{\mu\nu} $ is the curvature of the spinor-bundle, 
%
\begin{eqnarray}
{\cal F}_{\mu\nu}=2(\nabla_{[\mu}{\cal A}_{\nu]}+{\cal A}_{[\mu}  {\cal A}_{\nu]})\,. 
\end{eqnarray}
%
Using the identity (\ref{Curvatureid}), a straightforward computation brings the divergence of the Nester tensor into 
the following form, 
%
\begin{eqnarray}
\nabla_\nu N^{\mu\nu} & = & 2i {\overline {{\hat \nabla}_\rho \epsilon}}\gamma^{\mu\nu\rho}\hat \nabla_\nu \epsilon
-G^\mu{}_\nu V^\nu \nonumber \\
& & 
-\frac{i}{2}{\bar \epsilon} [{\bar {\cal F}}_{\nu\rho}\gamma^{\mu\nu\rho}+\gamma^{\mu\nu\rho} {\cal F}_{\nu\rho}
] \epsilon \nonumber \\
& & -i \bar \epsilon ({\bar {\cal A}}_\nu  \gamma^{\mu\nu\rho}- \gamma^{\mu\nu\rho}{\cal A}_\nu )
\hat \nabla_\rho \epsilon \nonumber \\
& & +i {\overline {{\hat \nabla}_\rho \epsilon}}({\bar {\cal A}}_\nu  \gamma^{\mu\nu\rho}- \gamma^{\mu\nu\rho}{\cal A}_\nu )
\epsilon\,, 
 \label{divnester}
\end{eqnarray}
%
where $V^\mu =i \bar \epsilon \gamma^\mu \epsilon$ is the future-directed timelike vector and 
${\bar {\cal F}}_{\mu\nu}:=\gamma^0 {\cal F}_{\mu\nu}^\dagger \gamma^0$.
Here 
we assume that $\epsilon$ satisfies the Dirac-Witten condition~\cite{Witten}
%
\begin{eqnarray}
\gamma^i \hat \nabla_i \epsilon=0\,, 
\label{DiracWitten}
\end{eqnarray}
%
where the Latin index $i$ stands for the spatial component. 
We suppose that this differential equation has no $L^2$-normalizable
zero-mode subjected to the boundary condition such that the 
spacetime energy-momentum is finite. This implies that the solution of (\ref{DiracWitten}) exists (this is indeed the case for 
asymptotically Minkowski/anti-deSitter spacetime in general relativity~\cite{Parker:1981uy}). 
It follows that the 0-th component of the first term on the right-hand side of (\ref{divnester}) is non-negative
$2i{\overline{{\hat\nabla}_\rho\epsilon}}\gamma^{0\nu\rho}\hat\nabla_\nu\epsilon=2g^{ij}
(\hat \nabla_i\epsilon)^\dagger \hat \nabla_j\epsilon \ge 0$. 
Our remaining task is to work out the condition under which 
the 0-th component of the residual terms is non-negative. 
In the simplest ${\cal A}_\mu=0$ case,  
the Einstein equations $G_{\mu\nu}=T_{\mu\nu}^{({\rm matter})}$
enable us to establish this claim, provided 
$T_{\mu\nu}^{({\rm matter})}$ satisfies the dominant energy condition. 
This proves that the ADM mass is positive in general relativity \cite{Witten}.

Next we are interested in the gravitational theory 
for which $\cal A_\mu $ is nonvanishing and    
the positivity of the mass holds.
Here we require that the 4th and 5th terms in the right-hand side of
Eq.~(\ref{divnester}) should vanish,  
since they have no sign control. This is realized if 
%
\begin{eqnarray}
{\bar {\cal A}}_\nu  \gamma^{\mu\nu\rho}=\gamma^{\mu\nu\rho}{\cal A}_\nu\,. \label{condA}
\end{eqnarray}
%
This condition is indeed satisfied for the Einstein-Maxwell system
with or without a negative cosmological constant~\cite{Gibbons:1982fy,Kostelecky:1995ei} . 
It deserves to notice that the Einstein-$\Lambda(>0)$ system
fails to fulfill Eq.~(\ref{condA})
\cite{Shiromizu:2001bg,Kastor:2002fu}.
This accords with our intuition because the
deSitter universe does not have a globally timelike Killing field.

When Eq.~(\ref{condA}) is satisfied, Eq.~(\ref{divnester}) is considerably simplified to  
%
\begin{align}
\nabla_\nu N^{\mu\nu}  =  2i {\overline{{\hat \nabla}_\rho \epsilon}}\gamma^{\mu\nu\rho}\hat \nabla_\nu \epsilon
-G^\mu{}_\nu V^\nu +S^\mu \,, 
\label{divnester1}
\end{align}
where 
\begin{align}
S^\mu := -i \bar \epsilon \gamma^{\mu\nu\rho}
\mathcal F_{\nu\rho}\epsilon \,.
\label{Smu}
\end{align}
Although the positivity of the 0-th component of 
$-G^\mu{}_\nu V^\nu +S^\mu $ depends sensitively on the theories under 
consideration,  the above formula (\ref{divnester1})
nevertheless shows the broad utility in a variety of gravitational theories due to its simplicity.

\section{Einstein-scalar system}
\label{sec:Eisnteinscalar}

In this section we apply the general formula obtained in the preceding section to 
specific gravitational theory. 
As a first stage of progress we focus on a quite simple, 
but still non-trivial system described by the action
%
\begin{eqnarray}
S=\frac{1}{2}\int \D ^4x {\sqrt {-g}}\left[
R+2K(\phi,X)+2{\cal L}_{\rm matter}\right]\,,\label{action}
\end{eqnarray}
%
where $\phi$ is a real scalar field, $X:=-(1/2)(\nabla \phi)^2$ and 
${\cal L}_{\rm matter}$ is the Lagrangian for ordinary matters which
decouples with the scalar field. This is the simplest 
case of the Horndeski theory (for example, see Refs. \cite{Horndeski, Kobayashi}). 
The gravitational field equation derived from the action is 
%
\begin{eqnarray}
G_{\mu\nu}= T^{(\phi )}_{\mu\nu }+ T_{\mu\nu}^{({\rm matter})}\,, \label{einstein}
\end{eqnarray}
with 
\begin{align}
 T^{(\phi)}_{\mu\nu }=K_X\nabla_\mu \phi \nabla_\nu \phi +Kg_{\mu\nu } \,,
\label{scalarEMtensor}
\end{align}
%
where $T_{\mu\nu}^{({\rm matter})}$ is the energy-momentum tensor 
for matters satisfying the dominant energy condition. Here and in what follows, 
the subscript denotes the partial derivative with respect to the 
corresponding argument, e.g., $K_X=\partial_X K(\phi,X)$. 
The evolution equation for the scalar field reads
%
\begin{eqnarray}
\nabla^\mu (K_X \nabla_\mu \phi)+K_\phi=0 \,. 
 \label{scalar}
\end{eqnarray}
%
It is obvious that the general form of $K$ does not 
ensure the mass positivity. For example, $K=-X$ 
gives rise to a ghost. In order to prevent this, we demand that 
the scalar field satisfies the null energy condition. This is tantamount to 
imposing a single condition  
\begin{align}
 K_X \ge 0 \,. \label{NEC}
\end{align}
We will use this equation later.

We now wish to examine the acceptable form of $\cal A_\mu $. 
Since the energy-momentum tensor (\ref{scalarEMtensor}) does not involve
the second derivative of $\phi$, $\cal A_\mu $ cannot have the derivative of 
$\phi$. Otherwise, the curvature ${\cal F}_{\mu\nu}$ 
in (\ref{Smu}) gives rise to a second derivative of $\phi$, 
which cannot be cancelled.   
Hence the simplest form of ${\cal A}_\mu$ is 
%
\begin{eqnarray}
{\cal A}_\mu = W(\phi)\gamma_\mu\,,  \label{connection}
\end{eqnarray}
for which 
${\cal F}_{\mu\nu} = 2W_\phi\nabla_{[\mu} \phi\gamma_{\nu]}+2W^2\gamma_{\mu\nu}$. 
This ${\cal A}_\mu$ satisfies the condition of Eq.~(\ref{condA})

Now the vector field $S^\mu $ given by~(\ref{Smu}) becomes 
%
\begin{align}
S^\mu =& -4i {\bar \epsilon}\gamma^{\mu\nu}\epsilon \nabla_\nu \phi W_\phi 
+12V^\mu  W^2 \nonumber \\
 = &  i {\overline {\delta \lambda}}\gamma^\mu \delta \lambda
+V^\nu 
\Bigl[ f^2 \nabla^\mu \phi \nabla_\nu \phi \nonumber \\
 & +\Bigl( -\frac{1}{2}f^2 (\nabla \phi)^2 -8f^{-2}W_\phi^2 +12W^2 \Bigr)\delta^\mu{}_\nu \Bigr]\,,
\label{Smu2} 
\end{align}
%
where we have defined $\delta \lambda$ as 
%
\begin{eqnarray}
\delta \lambda :=\frac{1}{{\sqrt {2}}}\left[f(\phi,X) \gamma^\mu \nabla_\mu \phi-4f^{-1}(\phi,X)W_\phi\right]
\epsilon\,.  \label{dilatino}
\end{eqnarray}
%
The second term in (\ref{dilatino}) has been chosen to cancel the 
$i\bar \epsilon \gamma^{\mu\nu }\epsilon $ term at the first equality of Eq.~(\ref{Smu2}). 
Comparing Eq.~(\ref{Smu2}) with (\ref{scalarEMtensor}), 
$S^\mu$ can be written into 
%
\begin{eqnarray}
S^\mu & =  i {\overline {\delta \lambda}}\gamma^\mu \delta \lambda +T^{(\phi)\mu }{}_\nu V^\nu \,, 
\label{Smu3}
\end{eqnarray}
%
provided $K$ satisfies 
%
\begin{eqnarray}
f^2=K_X(\phi,X) \label{K1}\,, 
\end{eqnarray}
%
and
%
\begin{eqnarray}
-\frac{1}{2}f^2 (\nabla \phi)^2 -8f^{-2}W_\phi^2 +12W^2=K\,.
 \label{K2}
\end{eqnarray}
%
If $K$ satisfies the above two relations, 
the use of Einstein's equations then allows us to find 
%
\begin{align} 
\nabla_\nu N^{\mu\nu}
= 2i {\overline {{\hat \nabla}_\rho \epsilon}}\gamma^{\mu\nu\rho}\hat \nabla_\nu \epsilon
-T^{\mu ({\rm matter})}_{~\nu}V^\nu + i {\overline {\delta \lambda}}\gamma^\mu \delta \lambda. 
\label{divnester2}
\end{align} 
%
Since the 0-th component of the last term is non-negative and 
$T^{({\rm matter})}_{\mu\nu}$ is supposed to satisfy 
the dominant energy condition, 
we can show the positivity of the mass.

We now look into in more detail the two conditions 
(\ref{K1}) and (\ref{K2}) required above. Obviously,  
Eq.~(\ref{K1}) requires that $K_X$ is non-negative, which is assured 
if we impose the null energy condition for the scalar field~(\ref{NEC}). 
Eqs. (\ref{K1}) and (\ref{K2}) lead to the following equation that 
$K$ should satisfy, 
%
\begin{eqnarray}
XK_X-K-\frac{8W_\phi^2}{K_X}=-12 W^2 (\phi) \,.
\label{K_diffeq}
\end{eqnarray}
%
From the above equation, we can obtain the integrability condition
\begin{align}
0= \partial_X\left( XK_X-K-\frac{8W_\phi ^2}{K_X}\right)=K_{XX}\left(X+\frac{8W_\phi^2}{K_X^2}\right) \,. 
\end{align} 
The general solutions to this equation fall into two families, 
\begin{align}
 {\rm (i)}~ K_{XX}=0 \,, \qquad
{\rm (ii)} ~~ XK_X^2 +8 W_\phi ^2 =0 \,. 
\label{Ksol}
\end{align}
The case (i) reduces to $K=X-U(\phi)$ (the coefficient of $X$ can be 
set to unity by absorbing into the rescaling of $\phi$), i.e., 
this corresponds to the canonical scalar field. In this case, Eq. (\ref{K_diffeq}) implies 
%
\begin{eqnarray}
U(\phi)=8W_\phi^2-12W^2 \,. \label{pot}
\end{eqnarray}
%
This recovers the result obtained in Refs.~\cite{Boucher,Townsend}. 
The potential $U(\phi)$ is now expressed in terms of a single 
superpotential $W(\phi)$.

The solution to case (ii) depends on the sign of $X$. 
For the $X>0$ case, one can see immediately that  
$K_X=W_\phi=0$ , thus $K=K(\phi)$ and $W={\rm constant}$. The field equation for 
$\phi$ shows $K={\rm constant}$. This is the case of the Einstein theory with a 
negative cosmological constant. 
Thus, there is no non-trivial scalar field in this case. 

For $X<0$, on the other hand, 
the differential equation in~(\ref{Ksol}) can be  integrated to give 
$K=4\sqrt 2 W_\phi (-X)^{1/2} +K_1 (\phi)$. Plugging this back into 
(\ref{K_diffeq}), we obtain $K_1=12 W^2(\phi)$, i.e.,
%
\begin{eqnarray}
K=4\sqrt 2 W_\phi (-X)^{1/2}+12W^2(\phi) \,. \label{case2-2}
\end{eqnarray}
%
Note that this excludes the  
homogeneous-isotropic solution with $\phi=\phi (t)$ because of $X=\dot \phi^2/2 >0$. 
This means that the case (ii) cannot be applied to a cosmological  
argument. 

Since the Lagrangian for $X<0$ takes a complicated form, 
it is far from obvious whether this theory admits asymptotically
Minkowski/anti-deSitter solutions. 
To see this more precisely, let us focus on the 
spherically symmetric static spacetimes, for which 
\begin{align}
 \D s^2 =-e^{2\nu (r)}\D t^2+e^{2\lambda (r)}\D r^2 +r^2 \D \Omega_2^2
 \,, \quad \phi=\phi(r) \,.
\end{align}
In this case we obtain 
\begin{align}
 \sqrt{-g} K&=4r^2 \sin\theta e^{\nu
 +\lambda}[-e^{-\lambda}W_\phi \phi'+3W^2]
\nonumber \\&\sim 
r^2 [-W'(r)(r/\ell)^n+W^2] \,,
\label{noncanK_asy}
\end{align}
where we have denoted 
$n=0$ for Minkowski  and $n=1$ for anti-deSitter with the curvature
radius $\ell$. We also assumed $\phi'(r)<0$ but this is not essential. 
When $n=0$ and $W_\phi \neq 0$, we have $W \sim 1/r$, while the $n=1$
case yields $W \sim 1/\ell+1/r$ to maintain the asymptotic conditions. 
Hence the non-canonical case also may allow asymptotically Minkowski/anti-deSitter
solutions, as far as the superpotential $W(r)$ displays the above
fall-off rate. 

It should be also emphasized, however,  that under these settings
$\phi(r)$-dependence disappears from Eq.~(\ref{noncanK_asy}), which now gives
a governing equation for $W= W(r)$ [of course, this gives an implicit equation for $\phi$
once we fix $W=W(\phi)$]. 
This is quite strange because the solution itself is insensitive to the
explicit form of $W=W(\phi)$ and $\phi=\phi (r)$ is undetermined.
This is due to the property that the theory (\ref{case2-2})
is invariant under the field re-definition  
\begin{align}
 \phi \to \Phi(\phi) \,.
\end{align} 
We will see in the next section that the 
supersymmetric solution illustrates this feature.

Finally let us consider the case in which the mass vanishes.
Eq.~(\ref{divnester2}) indicates that $\hat \nabla_i \epsilon=\delta \lambda=T_{\mu\nu}^{(\rm matter)}=0$
holds on $\Sigma$.  A slight deformation of the time slices implies 
$\hat \nabla_\mu \epsilon=\delta \lambda=T_{\mu\nu}^{(\rm matter)}=0$ holds. 
As shown in Ref. \cite{Townsend}, 
they imply that the spacetime is anti-deSitter/Minkowski and $\phi$ is constant if the 
spacetime is asymptotically globally anti-deSitter/Minkowski.\footnote{This is not the case for asymptotically 
locally anti-deSitter \cite{Horowitz}} This statement means that the ground state of the 
spacetime is maximally symmetric. This is a quite 
convincing and important consequence.

\section{Supersymmetric backgrounds}
\label{sec:SUSY}

Once an inequality has been proved for the system of Eq. (\ref{action}), 
one is next interested in the cases where the inequality is saturated. 
In the previous section we demonstrated that 
this is the case only for anti-deSitter/Minkowski, provided the 
spacetime globally approaches asymptotically the anti-deSitter/Minkowski spacetime.
In this section we relax the asymptotic boundary conditions and 
argue the interrelationships between the positive mass theorem 
and supersymmetry. 

In supergravity, the gravitational background is said to preserve 
supersymmetry or be in a BPS state if the variation of the fermionic configurations vanishes. 
This forces the spacetime to obey the first-order differential
equations, 
\begin{align}
 \hat \nabla_\mu \epsilon =0 \,, \qquad 
\delta \lambda= 0 \,. \label{Killingspinor}
\end{align}
In the present case, we have 
\begin{subequations}
\label{Killingspinor2}
\begin{align}
\left(\nabla_\mu +W(\phi)\gamma_\mu \right)\epsilon  &=0 \,, 
\label{gravi_KSeq}\\ 
\left(
\gamma^\mu \nabla_\mu \phi -\frac{4W_\phi}{K_X} 
\right) \epsilon &=0 
\label{dil_KSeq}\,, 
\end{align}
\end{subequations}
where $K$ is given by (\ref{pot}) or (\ref{case2-2}). 
The spinor $\epsilon $ obeying these equations is 
conventionally called a Killing spinor. This is because 
$V^\mu =i\bar \epsilon \gamma^\mu \epsilon $ is always a 
causal Killing vector when the theory 
can be embedded into the genuine supergravity theories (see~\cite{Maeda:2011sh}). In the generic gravitational theories, 
$V^\mu =i\bar \epsilon \gamma^\mu \epsilon $ becomes 
a Killing vector of BPS spacetime if the following condition holds
\begin{align}
 \bar{\cal A}_{(\mu }\gamma_{\nu ) }=\gamma_{(\mu }\cal A_{\nu )} \,. 
\label{Acond_Killing}
\end{align}
This condition may be also used to constrain the possible form of 
$\cal A_\mu $ and is fulfilled for the present model~(\ref{connection}). 
We shall refer respectively to  
(\ref{gravi_KSeq}) as a gravitino Killing spinor equation and (\ref{dil_KSeq})
as a dilatino Killing spinor equation, 
although they might not have a supergravity origin. 

It deserves to notice that the existence of the nontrivial BPS geometries
is not always assured~\cite{Nozawa:2010rf}.  
To check the consistency, we compute the integrability condition 
for the dilatino Killing spinor. 
Acting $\nabla^\nu\phi\nabla_\nu $ to Eq.~ (\ref{dil_KSeq}), we have 
\begin{align}
 \gamma^\mu \nabla_\mu X \epsilon = \biggl(& 
\frac{8XW_{\phi\phi}}{K_X}+\frac{4W_\phi}{K_X^2}\nabla^\mu \phi \nabla_\mu K_X
\nonumber \\
&+\frac{16WW_\phi ^2 }{K_X^2}+2 X W 
\biggl) \epsilon \,, 
\label{Xgamma_KS}
\end{align}
whereas acting $\gamma^\nu \nabla_\nu$ to $\delta \lambda=0$ yields 
\begin{align}
 0=&\gamma^\nu  \nabla_\nu \left[
\left(\sqrt{K_X}\gamma^\mu \nabla_\mu \phi 
-\frac{4W_\phi}{{\sqrt K_X}} \right)\epsilon 
\right] \nonumber \\
=& \biggl(\sqrt{K_X}\nabla^2\phi -\frac{16W_\phi W_{\phi\phi}}{K_X^{3/2}}\nonumber \\
&\quad +\frac{24WW_\phi}{\sqrt{K_X}}
+\frac{4W_\phi}{K_X^{3/2}}\gamma^\mu \nabla_\mu K_X \biggl)\epsilon \,. 
\label{int_dilatino}
\end{align}
Here we have used $\hat \nabla_\mu \epsilon =0$
and $\delta \lambda=0$.  Inserting (\ref{Xgamma_KS}) into the last term of~(\ref{int_dilatino}), 
we obtain 
\begin{align}
\frac{1}{\sqrt{K_X}} \left(K_X\nabla^2 \phi + \nabla^\mu \phi \nabla_\mu K_X +K_\phi \right) \epsilon =0 \,.
\end{align}
It turns out that the scalar-field equation~(\ref{scalar}) is automatically satisfied 
if there exists a spinor $\epsilon $ satisfying (\ref{gravi_KSeq})  and (\ref{dil_KSeq}).

Similarly, the integrability condition of the gravitino Killing spinor 
equation (\ref{Curvatureid}) implies 
\begin{align}
0 &= \gamma^\nu \hat \nabla_{[\mu }\hat \nabla_{\nu ]} \epsilon 
= -\frac{1}{4}\left(
R_{\mu\nu }\gamma^\nu -2 \gamma^\nu {\cal F}_{\mu\nu }
\right)\epsilon \,.
\end{align}
Substitution of the current model gives 
\begin{align}
E_{\mu\nu }\gamma^\nu \epsilon =0 \,, 
\end{align}
where 
\begin{align}
 E_{\mu\nu }:= R_{\mu\nu }-\left[
K_X\nabla_\mu \phi \nabla_\nu \phi +(XK_X-K)g_{\mu\nu }
\right] \,.
\end{align}
The vanishing $E_{\mu\nu }$ amounts to requiring Einstein's equations. 
When $V^\mu =i\bar \epsilon \gamma^\mu \epsilon $ is timelike, 
$E_{\mu\nu }=0$ follows automatically, whereas when  $V^\mu $ is null, 
except for the single component of Einstein's equations are 
satisfied. This is a common feature in supergravity and 
provides a consistency check~\cite{Maeda:2011sh}. 

We now classify all the BPS geometries satisfying
Eq.~(\ref{Killingspinor2}). To this end, we introduce the 
tensorial bilinears constructed from a Killing spinor~\cite{Caldarelli:2003pb}, 
\begin{align}
&E=\bar \epsilon \epsilon \,, \qquad B=i\bar \epsilon \gamma_5
 \epsilon\,, \qquad 
V_\mu =i\bar \epsilon\gamma_\mu \epsilon  \,,\nonumber \\
& 
U_\mu =i \bar \epsilon \gamma_5 \gamma_\mu \epsilon \,,
\qquad  \Phi_{\mu\nu }=i\bar \epsilon \gamma_{\mu\nu }\epsilon \,,
\end{align}
where $\gamma_5=i \gamma_{0123}$ is a chiral matrix ($\gamma_5^2=1$). 
In our convention, all the above tensorial fields are real.  Our strategy to classify BPS solutions is 
to constrain the possible form of metric by deriving algebraic and differential conditions for the bilinears. 
The algebraic relations of 
bilinears are obtained from Fierz identities and read 
\begin{align}
 &V\cdot V =-U\cdot U=-(E^2+B^2) \,, \qquad 
V\cdot U=0\,, 
\nonumber \\
& (E^2+B^2)\Phi_{\mu\nu }=2BV_{[\mu }U_{\nu]}
 -E\epsilon_{\mu\nu\rho\sigma }V^\rho U^{\sigma } \,. 
\label{Phi_Fierzid}
\end{align}
The gravitino Killing spinor equation now puts
differential constraints upon the bilinears, 
\begin{subequations}
\begin{align}
 \nabla_\mu E&=0 \,,\label{bilin_Eeq} \\
 \nabla_\mu  B&=- 2W U_\mu \,, \label{bilin_Beq}\\
 \nabla_\mu V_\nu &=2W\Phi_{\mu\nu }\,, \label{bilin_Veq}\\
 \nabla_\mu U_\nu &=-2WB g_{\mu\nu } \,, \label{bilin_Ueq}\\
\nabla_\mu \Phi_{\nu\rho }&=4W g_{\mu[\nu }V_{\rho] }\,.
\label{bilin_Phieq}
\end{align}
\end{subequations}
It follows that $V^\mu $ is a Killing vector and 
$U^{\mu } $ is a conformal Killing vector. 
Similarly, the dilatino equation gives 
\begin{subequations}
\begin{align}
 V^\mu \nabla_\mu \phi=0 \,,\label{BPhieq_Vphi} \\
\quad EW_\phi =0 \,, \quad  E\nabla_\mu \phi
 =0 \,,
\label{BPhieq_EWEphi} \\
U^{\mu } \nabla_\mu \phi -\frac{4W_\phi B}{K_X}=0 \,, \\
\Phi_{\mu\nu }\nabla^\nu \phi -\frac{4W_\phi}{K_X}V_\mu =0 \,, \\
B\nabla_\mu \phi -\frac{4W_\phi}{K_X}U_\mu =0 \,.
\label{BPhieq}
\end{align}
\end{subequations}
In this paper we are interested in supersymmetric solutions 
in the non-canonical case. 
As we will see below, we are able to find the general BPS metric 
in an explicit form utilizing algebraic and differential relations 
obtained above.

Due to (\ref{bilin_Eeq}), we can set $E=0$ or $E=1$.
In the latter case, Eq.~(\ref{BPhieq_EWEphi}) implies $\phi=W={\rm constant}$, which is 
not our main concern here. Hence we shall consider the $E=0$ case
in what follows.

Let us focus on the $B\ne 0$ case, for which 
$V^\mu $ is timelike and $U^\mu $ is spacelike.\footnote{
The null family $V^\mu V_\mu=E=B=0$ gives the plane-fronted wave. 
The explicit metric can be obtained in a similar fashion but we shall not attempt to do this
here. 
} 
Eqs. (\ref{Phi_Fierzid}) and  (\ref{bilin_Veq}) imply that $V^\mu $ is hypersurface-orthogonal, 
hence the spacetime is static.  It is therefore convenient to introduce a
local coordinate system for which $V^\mu=(\partial_t)^\mu $ 
with $V_\mu =-B^2\nabla_\mu t$, implying that $g_{\mu\nu }$ and $\phi$
are $t$-independent [see Eqs. (\ref{bilin_Veq}) and  (\ref{BPhieq_Vphi})].   
Eq. (\ref{bilin_Ueq}) implies that there exists a local function $z$ 
such that $U=\D z$.\footnote{
It is worthwhile to notice that
the fact that $U_\mu $ 
is  closed 
$\nabla_{[\mu}U_{\nu ]}=i\bar \epsilon \gamma_5\gamma_{[\mu }
\mathcal A_{\nu ]}\epsilon -i\bar \epsilon \bar{\cal A}_{[\mu }\gamma_{\nu]}\gamma_5\epsilon=0$
is a direct consequence of the condition~(\ref{condA}). 
} 
Now we can write the metric as 
\begin{align}
\D s^2 =-B^2\D t^2 +B^{-2}\left[\D z^2+B^4 e^{2 \psi}(\D x^2+\D y^2 )\right]\,,
\end{align}
where we have exploited the freedom of ($x, y$) to eliminate the cross
term $\D x \D z$ etc and fix to the conformally flat gauge. 
Eqs. (\ref{bilin_Beq}) and (\ref{BPhieq}) imply that 
$B$ and $\phi$ are functions of $z$ only, whereas 
(\ref{bilin_Ueq}) implies that $\psi $ is independent of $z$. 
As pointed out in~\cite{Klemm:2013eca}, the trace of Einstein's equation
assures the integrability condition of the Killing spinor. 
This yields $(\partial_x^2+\partial_y^2)\psi =0$, hence 
we have $\psi=F(\zeta )+\bar F(\bar \zeta) $, 
where $F$ is an analytic function of $\zeta:=x+iy$.
Using the remaining freedom $\zeta=x+iy \to f(\zeta )$ which keeps the 
conformally flat form,  we can set $ \psi =0$ without loss of generality. 
Thus the most general timelike class of BPS solution reads 
\begin{align}
 \D s^2 =B^{-2}(z)\D z^2 + B^2(z) (-\D t^2+ \D x^2+\D y^2)\,,
\end{align} 
with 
\begin{align}
 W=-\frac{1}{2}B'(z) \,, \qquad \phi=\phi(z)\,.
\end{align}
Here $B(z)$ and $\phi(z)$ are arbitrary functions.   
In this case one can integrate the Killing spinor equation directly
and find the solution, 
\begin{align}
 \epsilon =B^{1/2}(z)(1+\gamma_3) \epsilon_0 \,,
\end{align} 
where $\epsilon_0 $ is a constant spinor. 
Hence the metric preserves $1/2$-supersymmetry.
This is in contrast with the ${\cal N}=2$ gauged supergravity, 
for which the general BPS solutions preserve
only one quarter of the supersymmetry~\cite{Caldarelli:2003pb}.   
One can also verify that the maximally supersymmetric solution is exhausted by the 
anti-deSitter spacetime. 

As noticed in the previous section, $W=W(\phi)$ remains unfixed. 
In this sense, the non-canonical theory obtained here 
is not qualified as a well-defined gravitational theory.

\section{Multiple scalar fields}
\label{sec:multi}

We discuss in this section the 
 generalization of the argument in section III to the multiple scalar system. 
We shall consider the following action
\begin{align}
 S=\frac{1}{2}\int \D^4x\sqrt{-g}\left[
R+2K(\phi^I, X^{JK}) +2\mathcal L_{\rm matter}
\right] \,, \label{multi_action}
\end{align}
where $\phi^I$ ($I=1,...,N$) are real scalar fields and 
$X^{IJ}:=-(1/2)\nabla^\mu \phi^I\nabla_\mu\phi^J$. 
The stress-energy tensor derived from the action reads
\begin{align}
 T_{\mu\nu }^{(\phi)}=K_{IJ}\nabla_\mu \phi^I \nabla_\nu \phi^J +K g_{\mu\nu } \,, 
\end{align}
where we have denoted $K_{IJ}:=\partial K/\partial X^{IJ}$. 
We assume that the metric $K_{IJ}$ is invertible and positive-definite. 

Following the parallel argument given in section~\ref{sec:Eisnteinscalar}, let us discuss the condition under which the 
gravitational theory~(\ref{multi_action}) admits the positive mass. 
We define the super-covariant derivative as 
\begin{align}
 \hat \nabla_\mu \epsilon =
\left(\nabla_\mu +W(\phi^I)\gamma_\mu \right) \epsilon \,,
\label{multi_A}
\end{align}
and the variation of dilatino as 
\begin{align}
 \delta \lambda^I =\frac{1}{\sqrt 2} \left(\gamma^\mu\nabla_\mu \phi^I-4K^{IJ}\partial_J W \right)\epsilon \,, 
\end{align}
where $K^{IJ}$ is the inverse metric of $K_{IJ}$ and $\partial_IW:=\partial W/\partial \phi^I$. 
It follows that 
the vector field $S^\mu $ can be written into the desired form, 
\begin{align}
 S^\mu =T^{(\phi)\mu }{}_\nu V^\nu +i K_{IJ}  \overline{\delta \lambda ^I}\gamma^\mu \delta \lambda^J \,, 
\end{align}
provided that 
$K(\phi^I, X^{JK})$ satisfies the following equation, 
\begin{align}
X^{IJ}K_{IJ}-K-8K^{IJ}\partial_IW\partial_JW=-12W^2 \,. 
\end{align}
The same line of argument given in section~\ref{sec:Eisnteinscalar} enables 
us to find two classes of solutions to this differential equation. 
One is the canonical scalar fields, for which $K$ is given by 
\begin{align}
 K=G_{IJ}(\phi) X^{IJ}-U(\phi) \,, 
\end{align}
where $G_{IJ}$ describes the positive-definite moduli space metric 
and 
\begin{align}
 U(\phi)=8G^{IJ}\partial_I W \partial_J W-12 W^2 \,.
\end{align}
Here $G^{IJ}$ is the inverse of $G_{IJ}$. 
This recovers the result in~\cite{Boucher,Townsend}.

Another class is non-canonical and is reduced to  
\begin{align}
 K=4 \left(-2\partial_IW\partial_JW X^{IJ}\right)^{1/2}+12 W^2 \,. 
\end{align}
This is a simple generalization of non-canonical scalar field given in 
section~\ref{sec:Eisnteinscalar}.
This requires that $X^{IJ}$ is negative-definite and excludes the 
cosmological solution of the form $\phi^I=\phi^I(t)$.

\section{Summary and discussion}
\label{sec:summary}

Since the classical stability of spacetime is a fundamental issue,  
the positive mass may be a creditable guide for the construction of the
gravitational theory. Based on this belief, we explored in this paper 
whether the gravitational theories with a non-canonical scalar field allow the positive mass.  
In the case of the simplest Einstein-scalar system $R+2K(\phi, X)$ or
its obvious multiple extension,  
we found that possible theories are only two types if the 
gauge connection does not involve the derivatives of scalar fields
[see Eqs. (\ref{connection}) and (\ref{multi_A})]. One 
is the canonical scalar system with the specific potential of
Eq.~(\ref{pot}) \cite{Boucher, Townsend}.  
For the other case with $X=-(1/2)(\nabla \phi)^2>0$, 
it turns out that there is no non-trivial solution for $\phi$. This is a quite
striking result because this prohibits cosmological solutions.
In the $X<0$ case, we have a non-canonical theory compatible with
positive mass. But, this case also excludes the cosmological solution. 
We checked the consistency of Killing spinor equations and obtained the 
most general BPS metrics in the non-canonical theory. 
The BPS solutions illustrate that  
this theory exhibits a strange behavior in the sense that the function $\phi$ 
remains unfixed if it depends on a single valuable. 
With this issue in mind, we should discard this theory as a 
viable gravitational theory.

We obtained a single condition upon the gauge connection 
required for the positive mass~(\ref{condA})
{\it \`a la} Witten-Nester argument. 
This restricts the plausible connection $\cal A_\mu $, but we have not 
fully understood the underlying mathematical and physical implications. 
We remarked that in four dimensions the condition~(\ref{condA}) is equivalent to the fact that 
the one-form $U_\mu=i\bar \epsilon \gamma_5\gamma_\mu \epsilon $ is closed
when the Killing spinor equation is satisfied. 
This bilinear property has an obvious dimensional dependence. For $d=5$, 
the BPS metric can be written as a fibre over the (hyper-)K\"ahler
manifold with a (hyper-)K\"ahler form 
$\Phi_{\mu\nu }=i\bar\epsilon\gamma_{\mu\nu}\epsilon$~\cite{Gauntlett,Maeda:2011sh}, and Eq.~(\ref{condA})
directly implies that $\Phi_{\mu\nu }$ is closed. 
We believe that we can assign an underlying mathematical reason to 
these relations.  We would like to address this issue in a future publication.

There is a significant distance to the similar study 
for the nonlinear massive gravity and so on which have not 
been addressed so far from the viewpoint of positive mass.  
However, the current study will give us a hint because the Horndeski theory may be regarded as 
a proxy theory for them.  

There are many remaining issues. 
We employed the super-covariant derivative as (\ref{supercovariantderv}), which
yields a universal formula~(\ref{divnester1}) with  the extra condition (\ref{condA}). 
In this equation, the Einstein tensor appears explicitly, hence this formula is useful for
the Einstein frame. In other theories, the Jordan-frame formulation
may be more advantageous, which 
motivates us to seek the suitable Jordan-frame super-covariant
derivative. We leave this issue to a future study.   

In addition, 
we employed the connection $\cal A_\mu $ given by Eq.~(\ref{connection}). 
This is the minimal ansatz to have the positive-definite form of the mass, 
and other forms of the gauge connection  are conceivable. 
The Horndeski theory~\cite{Kobayashi} may be tractable in an analogous fashion. 
The strategy is as follows:
\begin{enumerate}
\item[(i)] Find the appropriate connection $\cal A_\mu $ satisfying
Eq.~(\ref{condA}).

\item[(ii)] Prove the positivity of $S^0$ given by~(\ref{Smu}) modulo
field equations and the dominant energy condition for 
additional matter fields. 
\end{enumerate}

In the process of (i), 
the condition~(\ref{Acond_Killing}) will be of help, since this 
should be satisfied for all supergravity theories. For example, 
Eq.~(\ref{condA}) is satisfied for 
${\cal A}_\mu=f(\phi,X)(\gamma_{\mu\nu}\nabla^\nu\phi-\nabla_\mu\phi)$
but Eq.~(\ref{Acond_Killing}) is not. 
In (ii), a feasible blueprint is to write $S^\mu $ into the form
\begin{align}
 S^\mu = S^\mu{}_\nu V^\nu +i\overline{\delta \lambda }\gamma^\mu \delta
 \lambda \,, 
\end{align} 
as we did in the main text. 
Here $S_{\mu\nu }$ should yield Einstein's equations
$G_{\mu\nu }=S_{\mu\nu }+T_{\mu\nu }^{(\rm matter)}$ 
and be derived from a covariant action. 
This will enable us to find the modified gravitational theories 
admitting the positive mass in a methodical manner. 
We hope to visit these issues in a future publication.

\bigskip
\begin{acknowledgments}
MN is grateful to Hideo Kodama for helpful conversations. 
TS thanks Tsutomu Kobayashi, Ryo Saito and Yuki Sakakihara for fruitful discussions in the early 
stage of this work.  
We thank Yukawa Institute, Kyoto University, where this work was initiated during the two workshops 
on ``Nonlinear massive gravity and its observational test''(YITP-T-12-04) 
and ``String theory, black holes and holography''(YITP-W-13-02). The work of MS is partly 
supported by  the MEXT Grant-in-Aid for Scientific Research on Innovative Areas No. 21111006. 
TS is supported by Grant-Aid for Scientific Research from Ministry of Education, Science,
Sports and Culture of Japan (Nos.~21244033 and 25610055).  
\end{acknowledgments}



\end{document}